\documentclass[conference]{llncs}
\usepackage{graphicx}
\usepackage[utf8]{inputenc}
\usepackage{paralist}
\usepackage{tikz}
\usepackage{rotating}
\usepackage{setspace}
\usepackage{mathtools}
\usepackage{bussproofs}
\usepackage{latexsym}
\DeclareGraphicsExtensions{.pdf,.png,.jpg}
\usepackage{amsmath, amsfonts, amssymb}

\begin{document}

\title{Resolution in Linguistic Propositional Logic based on Linear Symmetrical Hedge Algebra}

\author{Thi-Minh-Tam Nguyen\inst{1} 
			\and Viet-Trung Vu\inst{2} \and\\
			 The-Vinh Doan\inst{2}
			\and  Duc-Khanh Tran\inst{2}}
			
\institute{Faculty of Information Technology - Vinh University\\
			Email: nmtam@vinhuni.edu.vn
			\and School of Information and Communication Technology\\ 
			Hanoi University of Science and Technology\\
			Email: 
				trungvv91@gmail.com\\
				doanthevinh1991@gmail.com\\
				khanhtd@soict.hut.edu.vn}
\maketitle

\newtheorem{defi}{Definition}[section]
\newtheorem{pros}{Proposition}[section]
\newtheorem{examp}{Example}[section]

\begin{abstract} 
The paper introduces a propositional linguistic logic that serves as the basis for automated uncertain reasoning with linguistic information. First, we build a linguistic logic system with truth value domain based on a linear symmetrical hedge algebra. Then, we consider G\"{o}del's t-norm and t-conorm to define the logical connectives for our logic. Next, we present a resolution inference rule, in which two clauses having contradictory linguistic truth values can be resolved. We also give the concept of reliability in order to capture the approximative nature of the resolution inference rule. Finally, we propose a resolution procedure with the maximal reliability. 
\end{abstract}

\section{Introduction}

Automated reasoning is an approach to model human thinking. The resolution rule introduced by Robinson (1965) \cite{Lee71} marked an important point in studying automated reasoning. Resolution based on fuzzy set theory of Zadeh \cite{Zadeh65} has been studied to deal with uncertain information. In fuzzy logic, each clause has a membership function in $[0,1]$. Since then subtantial works \cite{Lee71,Shen89,Ebrahim01,Vojtas01,Mondal,WeigertTL93} have been done on the fuzzy resolution.  

In two-valued logic, each clause has a truth value $\mathsf{True}$ or $\mathsf{False}$. Therefore, the logical inference is absolutely accurate. However, in linguistic logic, each literal has a linguistic truth value such as $\mathsf{MoreTrue}$, $\mathsf{MoreFalse}$, $\mathsf{PossibleVeryTrue}$, $\mathsf{LessTrue},...$, where $\mathsf{True, False}$ are \textit{generators} and $\mathsf{More, PossibleVery, Less, \ldots}$ are strings of hedges which increase or decrease the semantic of generators. Thus the accuracy of logical inference is approximate.  For instance the two clauses $A^{\mathsf{True}} \vee B^\mathsf{MoreTrue}$ and $B^\mathsf{LessFalse} \vee C^{\mathsf{True}}$ can be resolved to obtain $A^{\mathsf{True}} \vee C^{\mathsf{True}}$.  However the literals $B^\mathsf{MoreTrue}$ and $B^\mathsf{LessFalse}$ are not totally contradictory, they are only contradictory at a certain degree. Consequently the resolution inference is only reliable at a certain degree. Therefore, when the inference is performed, the infered formula should be associated with a certain reliability.  Automated reasoning in linguistic logic has been attracting many researchers. Many works presented resolution algorithms in linguistic logics with truth value domain based on the implication lattice algebraic structures \cite{HeLXMR12,LP6,XuRKL00,XuRKL01,ZhongXLC12} or based on hedge algebra \cite{Programming,hedge-khangho,Phuong_Khang,LAPhuong_GMP}. 

Along the line of these research directions, we study automated reasoning based on resolution for linguistic propositional logic with truth value domain is taken from linear symmetrical hedge algebra. The syntax of linguistic propositional logic is constructed. To define the semantics of logical connectives we consider t-norm and t-conorm operators in fuzzy logic, specially t-norm and t-conorm operators of G\"{o}del and {\L}ukasiewicz. We show that logical connectives based on G\"{o}del connectives are more appropriate to construct logical connectives for our linguistic logic. A resolution rule and resolution procedure are given. The concept of reliability of inference is introduced in such a way that the reliability of the conclusion is smaller than or equal to the reliabilities of the premises. We also present a resolution procedure with maximal reliability and prove the soundness and completeness of the resolution procedure.

The paper is structured as follows:  Section \ref{sec:HA} introduces basic notions and results on linear symmetrical hedge algebras. Section \ref{sec:syntax-semantics} presents the syntax and semantics of our linguistic propositional logic with truth value domain based on linear symmetrical hedge algebra. The resolution rule and resolution procedure are introduced in Section \ref{sec:resolution}. Section \ref{sec:conclusion} concludes and draws possible future work.  Proofs of theorems, lemmas and proposition are presented in the Appendix.

\section{Preliminaries}
\label{sec:HA}

We recall only the most important definitions of hedge algebra for our work and refer the reader to \cite{Wechler,C2009,hedge-khangho} for further details.

We will be working with a class of abstract algebras of the form $AX=(X,G,H,\leq)$ where $X$ is a term set, $G$ is a set of generators, $H$ is a set of linguistic hedges or modifiers, and $\leq$ is a partial order on $X$. $AX$ is called a $\textit{hedge algebra}$ (HA) if it satisfies the following:
\begin{itemize}
\item Each hedge h is said to be positive w.r.t k, i.e. either kx $\geq$ x implies hkx $\geq$ kx or kx $\leq$ x implies hkx $\leq$ kx; similarly h is said to be negative w.r.t k, i.e. either kx $\geq$ x implies hkx $\leq$ kx or kx $\leq$ x implies hkx $\geq$ kx (for x $\in$ X);
\item If terms $u$ and $v$ are independent, then, for all $x \in H(u)$, we have $x \notin H(v)$. If $u$ and $v$ are incomparable, i.e. $u \not< v$ and $v \not< u$, then so are x and y, for every $x \in H(u)$ and $y \in H(v)$;
\item If $x \neq hx$, then $x \notin H(hx)$, and if $h \neq k$ and $hx \leq kx$, then $h'hx \leq k'kx$, for all $h, k, h', k' \in H$ and $x \in X$. Moreover, if $hx \neq kx$, then $hx$ and $kx$ are independent;
\item If $u \notin H(v)$ and $u \leq v (u \ge v)$, then $u \leq hv (u \ge hv)$ for any $h \in H$.
\end{itemize}

Let $AX=(X, G, H, \leq)$ where the set of generators $G$ contains exactly two comparable ones, denoted by $c^- < c^+$. For the variable Truth, we have $c^+ = True > c^- = False$. Such HAs are called \textit{symmetrical} ones. For symmetrical HAs, the set of hedges $H$ is decomposed into two disjoint subsets $H^+=\{h \in H| hc^+ > c^+\}$ and $H^-=\{h \in H| hc^+ < c^+\}$. Two hedges h and k are said to be converse if $\forall x \in X, hx \leq x$ iff $kx \geq x$, i.e., they are in different subsets; h and k are said to be compatible if $\forall x \in X, hx \leq x$ iff $kx \leq x$, i.e. they are in the same subset. Two hedges in each of the sets $H^+$ and $H^-$ may be comparable. Thus, $H^+$ and $H^-$ become posets.

A symmetrical HA $AX=(X, G = \{c^-, c^+\}, H, \leq)$ is called a \textit{linear symmetrical HA} (\textit{LSHA}, for short) if the set of hedges $H$ can be decomposed into $H^+=\{h \in H| hc^+ > c^+\}$ and $H^-=\{h \in H| hc^+ < c^+\}$, and $H^+$ and $H^-$ are linearly ordered.

Let $h_nh_{n-1}\ldots h_1u$, $k_mk_{m-1}\ldots k_1 u$ be the canonical presentations of values $x,y$ respectively. $x = y$ iff  $m = n$ and $h_{j}$ = $k_{j}$ for every $j \leq n$. If $x \neq y$ then there exists an  $j \leq min\{m,n\}+1$ (there is one convention is understood that if  $j=min\{m,n\}+1$, then $h_{j}=I$ where $j =n+1 \leq m$ or $k_{j}=I$ where $j=m+1 \leq n$) such that $h_{j'}$= $k_{j'}$ with all $j’<j$. Denote that $x_{<j} = h_{j-1}h_{j-2}\ldots h_1u=k_{j-1}k_{j-2}\ldots k_1u$, we have: 
$x<y$ iff $h_jx_{<j}<k_jx_{<j}$,
 $x>y$ iff $h_jx_{<j}>k_jx_{<j}$.

Let $x$ be an element of the hedge algebra $AX$ and the canonical representation of $x$ is $x = h_{n}...h_{1}a$ where $a \in \{c^+, c^-\}$. The contradictory element of $x$ is an element $y$ such that $y = h_{n}...h_{1}a'$ where $a' \in \{c^+, c^-\}$ and $a' \neq a$, denoted by $\overline{x}$. In  LSHA, every element $x \in X$ has a unique contradictory element in $X$.

It is useful to limit the set of values $X$ only consists of finite length elements. This is entirely suitable with the practical application in natural language, which does not consider infinite number of hedge of string.

From now on, we consider a LSHA $AX = (X, G, H, \leq, \neg, \lor,\land , \rightarrow)$ where
$G=\{\bot,\mathsf{False},\mathsf{W},\mathsf{True},\top\}$;
$\bot,\top$ and $\mathsf{W}$ are the least, the neutral, and the greatest elements of X, respectively; $\bot<\mathsf{False} < \mathsf{W} < \mathsf{True} < \top$.
\section{Propositional logic with truth value domain based on symmetrical linear hedge algebra}
\label{sec:syntax-semantics}

Below, we define the syntax and semantics of the linguistic propositional logic.
\begin{defi}
An alphabet consists of:
\begin{itemize}
\item constant symbols: $\mathsf{MoreTrue}, \mathsf{VeryFalse}, \bot, \top, ...$
\item propositional variables: $A, B, C, ...$
\item logical connectives: $\lor, \land, \rightarrow, \neg, \equiv$, and
\item auxiliary symbols: $\Box, (, ), ...$
\end{itemize}
\end{defi}
\begin{defi}
An atom is either a propositional variable or a constant symbol.
\end{defi}

\begin{defi}
Let A be an atom and $\alpha$ be a constant symbol. Then $A^\alpha$ is called a literal.
\end{defi}

\begin{defi}
Formulae are defined recursively as follows:
\begin{itemize}
	\item either a literal or a constant is a formula,
	\item if $P$ is a formula, then $\neg P$ is a formula, and
	\item if $P,Q$ are formulae, then $P \lor Q$, $P \land Q$, $P\rightarrow Q, P\leftrightarrow Q$  are formulae.
\end{itemize}
\end{defi}

\begin{defi}
A clause is a finite disjunction of literals, which is written as $ l_{1} \lor l_{2} \lor ... \lor l_{n}$, where $l_{i}$ is a literal. An empty clause is denoted by $\Box$.
\end{defi}

\begin{defi}
A formula F is said to be in conjunctive normal form (CNF) if it is a conjunction of clauses.
\end{defi}

In many-valued logic, sets of connectives called {\L}ukasiewicz, G\"{o}del, and product logic ones are often used. Each of the sets has a pair of residual t-norm and implicator.  However, we cannot use the product logic connectives when our truth values are linguistic.
 
We recall the operators t-norm(T) and t-conorm(S) on fuzzy logic. It is presented detailed in \cite{Graded,WeigertTL93}.

T-norm is a dyadic operator on the interval $[0,1]$: $T:[0,1]^2\longrightarrow [0,1]$, satisfying the following conditions:
			\begin{itemize}
				\item Commutativity: $T(x,y)=T(y,x)$,
				\item Associativity: $T(x,T(y,z))=T(T(x,y),z)$,
				\item Monotonicity: $T(x,y)\leq T(x,z)$ where $y\leq z$, and
				\item Boundary condition: $T(x,1)=x$.
			\end{itemize}
If T is a t-norm, then its dual t-conorm S is given by
$S(x,y)=1-T(1-x,1-y)$.

Let $K=\{n|n\in \mathbb{N} , n\leq N_0\}$. Extended T-norm is a dyadic operator $T_E : K^2 \longrightarrow K$ and satisfies the following conditions:
			\begin{itemize}
				\item Commutativity: $T_E(m,n)=T_E(n,m)$,
				\item Associativity: $T_E(m,T_E(n,p))=T_E(T_E(m,n),p)$,
				\item Monotonicity: $T_E(m,n)\leq T_E(m,p)$ where $n\leq p$, and
				\item Boundary condition: $T_E(n,N_0)=n$.
			\end{itemize}

                        The Extended T-conorm is given by: $S_E(m,n)=N_0-T_E(N_0-n,N_0-m)$. It is easy to prove that $S_E$ is commutative, associate, monotonous. The boundary condition of $S_E$ is: $S_E(0,n)=n$.

Two common pairs $(T,S)$ in fuzzy logic: G\"{o}del's$(T,S)$ and {\L}ukasiewicz's$(T,S)$ are defined as following:
\begin{itemize}
\item G\"{o}del:
\begin{itemize}
\item $T_{G}(m,n)=\min(m,n)$
\item $S_{G}(m,n)=\max(m,n)$
\end{itemize}
\item {\L}ukasiewicz:
\begin{itemize}
\item $T_{L}(m,n)=\max(0,m+n-N_0)$
\item $S_{L}(m,n)=\min(m+n, N_0)$
\end{itemize}
\end{itemize}

Given a SLHA AX , since all the values in AX are linearly ordered, we assume that they are $\bot=v_{0} \le v_{1} \le v_{2} \le \ldots \le v_{n} = \top$.

Clearly, the pair $(T, S)$ is determined only depending on $\max$ and $\min$ operators. Commonly, the truth functions for conjunctions and disjunctions are t-norms and t-conorms respectively.

\begin{examp} 
Consider a SLHA $AX=(X,\{\mathsf{True, False}\}, \{\mathsf{More, Less}\}, \leq)$ with $\mathsf{Less} < \mathsf{More}$. 
We assume the length of hedge string is limited at 1. Then $AX=\{v_{0}=\mathsf{MoreFalse}, v_{1}=\mathsf{False}, v_{2}=\mathsf{LessFalse}, v_{3}=\mathsf{LessTrue}, v_{4}=\mathsf{True}, v_{5}=\mathsf{MoreTrue}\}$.
We determine the truth value of logical connectives based on t-norm and t-conorm operators of G\"{o}del and {\L}ukasiewicz:
\begin{itemize}
\item G\"{o}del:
	\begin{itemize}
		\item $\mathsf{LessFalse} \vee \mathsf{False} = max \{\mathsf{LessFalse}, \mathsf{False}\} = \mathsf{LessFalse}$
		\item $\mathsf{MoreTrue} \wedge \mathsf{True} = min \{ \mathsf{MoreTrue}, \mathsf{True}\}= \mathsf{True}$
		
	\end{itemize}
\item {\L}ukasiewicz:
	\begin{itemize}
		\item $\mathsf{LessFalse} \vee \mathsf{False} = \mathsf{LessTrue}$
		\item $\mathsf{MoreTrue} \wedge \mathsf{True} = \mathsf{LessFalse}$
	\end{itemize}
\end{itemize}
\end{examp}

In fact, if the same clause has two truth values $\mathsf{LessFalse}$
or $\mathsf{False}$, then it should get the value
$\mathsf{LessFalse}$. In the case of two truth values are
$\mathsf{MoreTrue}$ and $\mathsf{True}$ then it should get the value
$\mathsf{True}$. We can see that the logical connectives based on
G\"{o}del's t-norm and t-conorm operators are more suitable in the
resolution framework than those based on {\L}ukasiewicz's. In this
paper we will define logical connectives using G\"{o}del's t-norm and
t-conorm operators.

\begin{defi}
\label{logical_connections}
Let $S$ be a linguistic truth domain, which is a SLHA $AX=(X, G,
H,\leq)$, where $G = \{\top, \mathsf{True}, \mathsf{W},
\mathsf{False}, \bot \}$. The logical
connectives $\wedge$ (respectively $\vee$) over the set $S$ are
defined to be G\"{o}del's t-norm (respectively t-conorm), and
furthermore to satisfy the following:
\begin{itemize}
	\item $\neg \alpha = \overline \alpha$.
	\item $\alpha \rightarrow \beta = (\neg \alpha)\vee \beta$.
\end{itemize}
where $\alpha,\beta \in S$.
\end{defi}

\begin{pros}
\label{pros:property_connection}
Let $S$ be a linguistic truth domain, which is a SLHA $AX=(X,\{\top, \mathsf{True}, \mathsf{W}, \mathsf{False}, \bot \}, H,\leq)$; $\alpha,\beta, \gamma \in X$, we have:
\begin{itemize}
	\item Double negation:
		\begin{itemize}	
			\item $\neg \neg \alpha = \alpha$
		\end{itemize}
	\item Commutative:
		\begin{itemize}
			\item $\alpha \wedge \beta = \beta \wedge \alpha$
			\item $\alpha \vee \beta = \beta \vee \alpha$
		\end{itemize}
	\item Associative:
		\begin{itemize}
			\item $(\alpha \wedge \beta)\wedge \gamma = \alpha\wedge(\beta\wedge\gamma)$
			\item $(\alpha \vee \beta)\vee \gamma = \alpha\vee(\beta\vee\gamma)$
		\end{itemize}
	\item Distributive:
		\begin{itemize}
			\item $\alpha\wedge(\beta\vee\gamma)=(\alpha\wedge\beta)\vee(\alpha\wedge\gamma)$
			\item $\alpha\vee(\beta\wedge\gamma)=(\alpha\vee\beta)\wedge(\alpha\vee\gamma)$
		\end{itemize}
		\end{itemize}
		\end{pros}

\begin{defi}
An interpretation consists of the followings:
\begin{itemize}
\item a linguistic truth domain, which is a SLHA
  $AX=(X,G,H,\leq)$, where the set of generators $G = \{
  \top, \mathsf{True}, \mathsf{W}, \mathsf{False}, \bot \}$,
\item for each constant in the alphabet, the assignment of an element in $X$,
\item for each formula, the assignment of a mapping from $X$ to $X$.
\end{itemize}
\end{defi}

\begin{defi}
  Let $I$ be an interpretation and $A$ be an atom such that $I(A) =\alpha_1$. Then the truth value of a literal $A^{\alpha_2}$ under the interpretation $I$ is determined uniquely as follows:
\begin{itemize}
\item $I(A^{\alpha_2}) = \alpha_1 \land \alpha_2$ if $\alpha_1,\alpha_2 > \mathsf{W}$,
\item $I(A^{\alpha_2}) = \neg (\alpha_1 \lor \alpha_2)$ if $\alpha_1,\alpha_2 \leq \mathsf{W}$,
\item $I(A^{\alpha_2}) = (\neg \alpha_1) \lor \alpha_2$ if $\alpha_1 >
  \mathsf{W}, \alpha_2 \leq \mathsf{W}$, and
\item $I(A^{\alpha_2}) = \alpha_1 \lor (\neg \alpha_2)$ if $\alpha_1 \leq \mathsf{W}, \alpha_2 > \mathsf{W}$.
\end{itemize}
\end{defi}

\begin{defi}
The truth value of formulae under an interpretation is determined recursively as follows:
\begin{itemize}
\item $I(P\lor Q)=I(P)\lor I(Q)$,
\item $I(P\land Q)=I(P)\land I(Q)$,
\item $I(\neg P)=\neg I(P)$,
\item $I(P\rightarrow Q)=I(P)\rightarrow I(Q)$
\end{itemize}
\end{defi}

The following result follows from the properties of the $\land$ and $\lor$ operators.
\begin{pros}
\label{theo:formula}
Let $A$, $B$ and $C$ are formulae, and I be an arbitrary interpretation. Then,
\begin{itemize} 
\item Commutative: 
\begin{itemize}
	\item $I(A \lor B) = I(B \lor A)$
	\item $I(A \land B) = I(B \land A)$
\end{itemize}
\item Associative: 
\begin{itemize}
	\item $I((A\lor B)\lor C)=I(A\lor(B\lor C))$
	\item $I((A\land B)\land C)=I(A\land(B\land C))$
\end{itemize}
\item Distributive: 
\begin{itemize}
	\item $I(A\lor (B\land C))=I((A\lor B)\land(A\lor C))$
	\item $I(A\land (B\lor C))=I((A\land B)\lor(A\land C))$
\end{itemize}
\end{itemize}
\end{pros}

\begin{defi}
Let $F$ be a formula and $I$ be an interpretation. Then
\begin{itemize}
\item $F$ is said to be true under interpretation $I$ iff $I(F) \geq \mathsf{W}$, $I$ is also said to satisfy formula $F$, $F$ is said to be satisfiable iff there is an interpretation $I$ such that $I$ satisfies $F$, $F$ is said to be tautology iff it is satisfied by all interpretations;
\item $F$ is said to be false under interpretation $I$ iff $I(F) \leq \mathsf{W}$, $I$ is also said to falsify formula $F$, $F$ is said to be unsatisfiable iff it is falsified by all interpretations.
\end{itemize}
\end{defi}

\begin{defi}
\label{theo:consistent}
Formula $B$ is said to be a logical consequence of formula $A$,
denoted by $A \models B$, if for all interpretation $I$, $I(A) >
\mathsf{W}$ implies that $I(B) > \mathsf{W}$.
\end{defi}

\begin{pros}
\label{pros:consequen}
Let $A$ and $B$ be formulae. Then,
$A \models B$ iff $\models (A \rightarrow B)$.
\end{pros}

\begin{defi}
\label{defi:equivalence}
Two formulae $A$ and $B$ are logically equivalent, denoted by $A
\equiv B$, if and only if $A \models B$ and $B \models A$.
\end{defi}

\begin{pros}
\label{pros:equivalence}
Let $A,B$ and $C$ be formulae. Then the following properties hold:

\begin{itemize}
\item Idempotency: 
\begin{itemize}
\item $A\lor A \equiv A$
\item $A\land A \equiv A$
\end{itemize}

\item Implication: 
\begin{itemize}
\item $A\rightarrow B \equiv (\neg A)\lor B$
\item $(A\equiv B) \equiv (A\rightarrow B)\land (B\rightarrow A)$
\end{itemize}

\item Double negation: 
\begin{itemize}
\item $\neg \neg A \equiv A$
\end{itemize}

\item De Morgan:
\begin{itemize}
\item $\neg(A\lor B)\equiv (\neg A)\land(\neg B)$
\item $\neg(A\land B)\equiv (\neg A)\lor(\neg B)$
\end{itemize}

\item Commutativity:
\begin{itemize}
\item $A\lor B\equiv B\lor A$
\item $A\land B\equiv B\land A$
\end{itemize}
 
\item Associativity:
\begin{itemize}
\item $A\lor(B\lor C)\equiv (A\lor B)\lor C$
\item $A\land(B\land C)\equiv (A\land B)\land C$
\end{itemize}
 
\item Distributivity:
\begin{itemize}
\item $A\lor(B\land C)\equiv (A\lor B)\land(A\lor C)$
\item $A\land(B\lor C)\equiv (A\land B)\lor(A\land C)$
\end{itemize}
 
\end{itemize}
\end{pros}
We will be working with resolution as the inference system of our logic. Therefore formulae need to be converted into conjunctive normal form. The equivalence properties in Proposition~\ref{pros:equivalence} ensure that the transformation is always feasible.
\begin{theorem}
Let $F$ be a formula of arbitrary form. Then $F$ can be converted into an equivalent formula in conjunctive normal form.
\end{theorem}

\section{Resolution}
\label{sec:resolution}
In the previous section, we have described the syntax and semantics of our linguistic logic. In this section, we present the resolution inference rule and the resolution procedure for our logic.

\begin{defi}
The clause $C$ with reliability $\alpha$ is the pair $(C,\alpha)$ where $C$ is a clause and $\alpha$ is an element of SLHA $AX$ such that $\alpha > \mathsf{W}$. The same clauses with different reliabilities are called variants. That is $(C,\alpha)$ and $(C,\alpha')$ are called variants of each other.
\end{defi}

For a set of $n$ clauses $S = \{ C_1, C_2, ... , C_n \}$, where each $C_i$ has a reliability  $\alpha_i$, then the reliability $\alpha$ of $S$ is defined as:
$\alpha = \alpha_1 \land \alpha_2 \land ... \land \alpha_n$.

An inference rule $\mathsf{R}$ with the reliability $\alpha$ is represented as:
\[
\frac{(C_1,\alpha_1), (C_2,\alpha_2), \ldots ,(C_n,\alpha_n)}{(C,\alpha)}
\]

We call $\alpha$ the reliability of $\mathsf{R}$, provided that
$\alpha \leq \alpha_i$ for $i=1..n$.

\begin{defi}
\label{defi:resolution-rule}
The fuzzy linguistic resolution rule is defined as follows:
$$\frac{(A^a \lor B^{b_1},\alpha_1), (B^{b_2} \lor C^c, \alpha_2)}{(A^a \lor C^c, \alpha_3)}$$
where $b_1,b_2$ and $\alpha_3$ satisfy the following conditions:
\begin{displaymath}
\left \{
\begin{array}{l}
b_1 \land b_2 \leq \mathsf{W} ,\\
b_1 \lor b_2 > \mathsf{W} ,\\
\alpha_3 = f(\alpha_1, \alpha_2, b_1, b_2) 
\end{array}
\right.
\end{displaymath}
with $f$ is a function ensuring that $\alpha_3 \le \alpha_1$ and
$\alpha_3 \le \alpha_2$.
\end{defi}

$\alpha_3$ is defined so as to be smaller or equal to both $\alpha_1$ and $\alpha_2$. In fact, the obtained clause is less reliable than original clauses. The function $f$ is defined as following:
\begin{align}
\label{formula:combine-reliability}
\alpha_3 = f(\alpha_1, \alpha_2, b_1, b_2) = \alpha_1 \land \alpha_2 \land (\neg(b_1 \land b_2)) \land (b_1\lor b_2)
\end{align}
Obviously, $\alpha_1,\alpha_2 \ge W$, and $\alpha_3$ depends on $b_1,b_2$. Additionally, $b_1 \land b_2 \leq \mathsf{W}$ implies $\neg (b_1 \land b_2) >\mathsf{W}$. Moreover, $(b_1 \lor b_2) > \mathsf{W}$.  Then, by Formula (\ref{formula:combine-reliability}), we have $\alpha_3 >\mathsf{W}$.

\begin{lemma}
\label{lema:soundness_rule}
The fuzzy linguistic resolution rule~\ref{defi:resolution-rule} is sound.
\end{lemma}
 
We define a fuzzy linguistic resolution \emph{derivation} as a sequence of the form $S_0, \ldots, S_i, \ldots$, where:
\begin{itemize}
\item each $S_i$ is a set of clauses with a reliability, and
\item $S_{i+1}$ is obtained by adding the conclusion of a fuzzy
  linguistic resolution inference with premises from $S_i$, that is
  $S_{i+1} = S_i \cup \{(C,\alpha)\}$, where $(C,\alpha)$ is the
  conclusion of the fuzzy linguistic resolution
  $$\frac{(C_1,\alpha_1), (C_2, \alpha_2)}{(C,\alpha)},$$
  and $(C_1,\alpha_1), (C_2, \alpha_2) \in S_i$.
\end{itemize}

A \emph{resolution proof} of a clause $C$ from a set of clauses $S$ consists of repeated application of the resolution rule to derive the clause $C$ from the set $S$. If $C$ is the empty clause then the proof is called a \emph{resolution refutation}. We will represent resolution proofs as \emph{resolution trees}. Each tree node is labeled with a clause. There must be a single node that has no child node, labeled with the conclusion clause, we call it the root node. All nodes with no parent node are labeled with clauses from the initial set $S$. All other nodes must have two parents and are labeled with a clause $C$
such that $$\frac{C_1,C_2}{C}$$ where $C_1, C_2$ are the labels of the two parent nodes. If $\mathsf{RT}$ is a resolution tree
representing the proof of a clause with reliability $(C, \alpha)$,
then we say that $\mathsf{RT}$ has the reliability $\alpha$.

Different resolution proofs may give the same the conclusion clause with different reliabilities. The following example illustrate this.

\begin{examp}
\label{eg:RT}
Let $AX = (X, G, H, \leq, \neg,\lor,\land , \rightarrow)$ be a SRHA where $G=\{\bot,\mathsf{False},$ $\mathsf{W},\mathsf{True},\top\}$,
 $\bot,\mathsf{W},\top$ are the smallest, neutral,biggest elements respectively, and
 $\bot<\mathsf{False} < \mathsf{W} < \mathsf{True} < \top$;$H^+ =$ $\{\mathsf{V}$,$\mathsf{M}\}$ and $H^- = \{\mathsf{P, L}\}$ (V=Very, M=More, P=Possible, L=Less); 
Consider the following set of clauses:
\begin{enumerate}
\item $A^\mathsf{MFalse} \lor B^\mathsf{False} \lor C^\mathsf{VMTrue}$
\item $B^\mathsf{LTrue} \lor C^\mathsf{PTrue}$
\item $A^\mathsf{PTrue}$
\item  $B^\mathsf{VTrue}$
\item $C^\mathsf{VFalse}$
\end{enumerate}
At the beginning, each clause is assigned to the highest reliability $\top$. We have:

\begin{prooftree}
\label{eg:RT2}
\AxiomC{($A^\mathsf{PTrue},\top$)}
\AxiomC{($A^\mathsf{MFalse} \lor B^\mathsf{False} \lor C^\mathsf{VMTrue}, \top$)}
\kernHyps{-.5in}\insertBetweenHyps{\hskip-.10in}
\BinaryInfC{$(B^\mathsf{False} \lor C^\mathsf{VMTrue}, \mathsf{PTrue})$}
\AxiomC{($B^\mathsf{VTrue}, \top$)}
\kernHyps{-.5in}\insertBetweenHyps{\hskip-.10in}
\BinaryInfC{$(C^\mathsf{VMTrue}, \mathsf{PTrue})$}
\AxiomC{($C^\mathsf{VFalse}, \top$)}
\BinaryInfC{$(\Box, \mathsf{PTrue})$}
\end{prooftree}

\begin{prooftree}
\label{eg:RT3}

\AxiomC{($A^\mathsf{MFalse} \lor B^\mathsf{False} \lor C^\mathsf{VMTrue}, \top$)}
\AxiomC{($B^\mathsf{LTrue} \lor C^\mathsf{PTrue},\top$)}
\kernHyps{-.5in}\insertBetweenHyps{\hskip-.10in}
\BinaryInfC{$(A^\mathsf{MFalse} \lor C^\mathsf{VMTrue}, \mathsf{LTrue})$}
\AxiomC{$(A^\mathsf{PTrue},\top$)}
\kernHyps{-.5in}\insertBetweenHyps{\hskip-.20in}
\BinaryInfC{$(C^\mathsf{VMTrue}, \mathsf{LTrue})$}
\AxiomC{($C^\mathsf{VFalse}, \top$)}
\BinaryInfC{$(\Box, \mathsf{LTrue}) \fCenter$}
\end{prooftree}
\end{examp}

Since different proofs of the same clause may have different reliabilities, it is natural to study how to design a resolution procedure with the best reliability. Below we present such a  procedure.

We say that a set of clauses $S$ is \emph{saturated} iff
for every fuzzy linguistic resolution inference with premises in $S$, the conclusion of this inference is a variant with smaller or equal reliability of some clauses in $S$. That is for every fuzzy linguistic resolution inference
$$\frac{(C_1,\alpha_1), (C_2, \alpha_2)}{(C,\alpha)}$$ 
where $(C_1,\alpha_1), (C_2, \alpha_2) \in S$, there is some clause
$(C,\alpha') \in S$ such that $\alpha \leq \alpha'$.

We introduce a resolution strategy, called \emph{$\alpha$-strategy},
which guarantees that the resolution proof of each clause has the maximal reliability. An $\alpha$-strategy derivation is a sequence of the form
$S_0, \ldots, S_i, \ldots$, where 
\begin{itemize}
\item each $S_i$ is a set of clauses with reliability, and
\item $S_{i+1}$ is obtained by adding the conclusion of a fuzzy
  linguistic resolution inference with premises with maximal
  reliabilities from $S_i$, that is $S_{i+1} = S_i \cup
  \{(C,\alpha)\}$, where $(C,\alpha)$ is the conclusion of the fuzzy
  linguistic resolution inference
  $$\frac{(C_1,\alpha_1), (C_2, \alpha_2)}{(C,\alpha)}$$ 
  $(C_1,\alpha_1), (C_2, \alpha_2) \in S_i$ and there are not any
  clauses with reliability $(C_1,\alpha_1')$, $(C_2, \alpha_2') \in S_i$
  such that $\alpha_1' > \alpha_1$ and $\alpha_2' > \alpha_2$, or
\item $S_{i+1}$ is obtained by removing a variant with smaller
  reliability, that is $S_{i+1} = S_i \setminus \{(C,\alpha)\}$ where
  $(C,\alpha) \in S_i$ and there is some $(C,\alpha') \in S_i$ such
  that $\alpha < \alpha'$.
\end{itemize}
Define \emph{the limit of a derivation} $S_0, \ldots, S_i, \ldots$
  $$S_\infty = \bigcup_{i \geq 0} \bigcap_{j \geq i} S_j$$

The following result establishes the soundness and completeness of the resolution procedure.
\begin{theorem}
\label{theo:derivation}
  Let $S_0, \ldots, S_i, \ldots$ be a fuzzy linguistic resolution
  $\alpha$-strategy derivation. $S_n$ contains the empty clause iff $S_0$ is unsatisfiable (for some $n=0, 1, \ldots$).
\end{theorem}

\begin{lemma}
\label{lemma:RT2}
Consider the following resolution inferences:
$$ \frac{(A^a \lor B^{b_1},\alpha),(B^{b_2} \lor C^c,\beta)}{(A^a \lor C^c,\gamma)}$$
$$ \frac{(A^a \lor B^{b_1},\alpha),(B^{b_2} \lor C^c,\beta')}{(A^a \lor C^c,\gamma')}$$
Then, $\beta' > \beta$ implies $\gamma' \geq \gamma$.
\end{lemma}

\begin{lemma}
\label{lemma:saturated}
  Let $S_0, \ldots, S_i, \ldots$ be a fuzzy linguistic resolution
  $\alpha$-strategy derivation, and $S_\infty$ be the the limit of the
  derivation.  Then $S_\infty$ is saturated.
\end{lemma}

\begin{theorem}
\label{theo:max_reli}
  Let $S_0, \ldots, S_i, \ldots$ be a fuzzy linguistic resolution
  $\alpha$-strategy derivation, and $S_\infty$ be the the limit of the derivation.  Then for each clause $(C, \alpha)$ in $S_\infty$, there is not any other resolution proof of the clause $(C, \alpha')$ from
  $S_0$ such that $\alpha' > \alpha$.
\end{theorem}

\begin{examp}
Consider again Example \ref{eg:RT}. Applying the $\alpha$-strategy we
get the following saturated set of clauses
\begin{enumerate}
\item $(A^\mathsf{MFalse} \lor B^\mathsf{False} \lor C^\mathsf{VMTrue},\top)$
\item $(B^\mathsf{LTrue} \lor C^\mathsf{PTrue},\top)$
\item $(A^\mathsf{PTrue},\top)$
\item $(B^\mathsf{VTrue},\top)$
\item $(C^\mathsf{VFalse},\top)$
\item $(B^\mathsf{False} \lor C^\mathsf{VMTrue},\mathsf{PTrue})$
\item $(A^\mathsf{MFalse} \lor C^\mathsf{VMTrue},\mathsf{True})$
\item $(A^\mathsf{MFalse} \lor C^\mathsf{VMTrue},\mathsf{LTrue})$
\item $(A^\mathsf{MFalse} \lor B^\mathsf{False},\mathsf{VTrue})$
\item $(C^\mathsf{VMTrue},\mathsf{PTrue})$
\item $(A^\mathsf{MFalse},\mathsf{True})$
\item $(\Box, \mathsf{PTrue})$
\end{enumerate}
The initial set of clauses is unsatisfiable, and the resolution
futation is the following
\begin{prooftree}
\AxiomC{$(A^\mathsf{MFalse} \lor B^\mathsf{False} \lor C^\mathsf{VMTrue},\top)$}
\AxiomC{$(C^\mathsf{VFalse},\top)$}
\kernHyps{-.5in}\insertBetweenHyps{\hskip-.10in}
\BinaryInfC{$(A^\mathsf{MFalse} \lor B^\mathsf{False},\mathsf{VTrue})$}
\AxiomC{$(B^\mathsf{VTrue},\top)$}
\kernHyps{-.5in}\insertBetweenHyps{\hskip-.10in}
\BinaryInfC{$(A^\mathsf{MFalse},\mathsf{True})$}
\AxiomC{$(A^\mathsf{PTrue},\top)$}
		\BinaryInfC{$(\Box,\mathsf{PTrue})$}
\end{prooftree}
\end{examp}

\section{Conclusion}
\label{sec:conclusion}

We have presented a linguistic logic system with the basic components: syntax, semantics and inference. The syntax have been defined as usual. To define the semantics, the truth value domain have been taken from linear symmetrical hedge algebra and logical connectives have been defined based on G\"{o}del's t-norm and t-conorm. We have also introduced an inference rule associated with a reliability which guarantees that the reliability of the inferred clause is less than or equal to reliaility of the premise clauses. Moreover, we have given a resolution procedure which ensures that the proof of clauses has the maximal reliability. The soundness and completeness of the resolution procedure are also proved. The proofs of the theorems, proposititions and lemmas are omitted due to lack of space. They can be found in the full paper at http://arxiv.org/submit/769464/view. There are several lines of future works. It would be natural to consider the linguistic first order logic in the same settings as our logic here. It would be worth investivating how to extend our result to other hedge algebra structures and to other automated reasong methods.

\bibliographystyle{plain} 
\bibliography{References}

\newpage
\section*{Appendix}
\begin{appendix}
\section{Proof of Propositition~\ref{pros:consequen}}
\textbf{Proposition~\ref{pros:consequen}. } Let $A$ and $B$ be formulae. Then, $A \models B$ iff $\models (A \rightarrow B)$.

\begin{proof}
  Assume that $A \models B$, for any interpretation $I$, then if $I(A)
  < \mathsf{W}$, $I(\neg A) > \mathsf{W}$; otherwise if $I(A) >
  \mathsf{W}$, we recall that $A \models B$, so $I(B) >
  \mathsf{W}$. Hence, $I(A \rightarrow B) = \neg I(A) \lor I(B) >
  \mathsf{W}$. In other words, $\models (A \rightarrow
  B)$. Conversely, by a similar way, we can also show that $\models (A
  \rightarrow B)$ implies $A \models B$.
\end{proof}

\section{Proof of Lemma~\ref{lema:soundness_rule}}
\textbf{Lemma~\ref{lema:soundness_rule}. } The fuzzy linguistic resolution rule~\ref{defi:resolution-rule} is sound.
\label{proof_sound}
\begin{proof}
  We need to prove that for any interpretation $I$, if $I((A^a \lor
  B^{b_1}) \land (B^{b_2} \lor C^c)) > \mathsf{W}$ then  $I (A^a
  \lor C^c) > \mathsf{W}$.  We have that
\begin{align*}
&I((A^a \lor B^{b_1}) \land (B^{b_2} \lor C^c)) \\ 
&= I((A^a \land B^{b_2}) \lor (A^a \land C^c) \lor (B^{b_1} \land B^{b_2}) \lor (B^{b_1} \land C^c)) \\
&= I(A^a \land B^{b_2}) \lor I(A^a \land C^c) \lor I(B^{b_1} \land B^{b_2}) \lor I(B^{b_1} \land C^c) 
\end{align*}
It is easy to show that:
\begin{itemize}
\item $I(A^a \land B^{b_2}) \leq I(A^a) \leq I(A^a \lor C^c)$,
\item $I(A^a \land C^c) \leq I(A^a) \leq I(A^a \lor C^c)$,
\item $I(B^{b_1} \land C^c) \leq I(C^c) \leq I(A^a \lor C^c)$, and
\item $I(B^{b_1} \land B^{b_2}) \leq \mathsf{W}$
\end{itemize}
So, if $I(A^a \lor C^c) \leq \mathsf{W}$ then we must have that 
$$ I(A^a \land B^{b_2}) \lor I(A^a \land C^c) \lor I(B^{b_1} \land
B^{b_2}) \lor I(B^{b_1} \land C^c) \leq \mathsf{W}$$ which contradicts with
the initial assumption. This completes the proof of the theorem.
\end{proof}
\section{Proof of Theorem~\ref{theo:derivation}}
\textbf{Theorem~\ref{theo:derivation}. }
Let $S_0, \ldots, S_i, \ldots$ be a fuzzy linguistic resolution $\alpha$-strategy derivation. $S_n$ contains the empty clause iff $S_0$ is unsatisfiable (for some $n=0, 1, \ldots$).
\label{proof_derivation}
\begin{proof} First, we prove if $S_n$ contains the empty clause then $S_0$ is unstatisfialbe.

If $S_n$ contains the empty clause, then $S_0$ is false under any interpretation. By Lemma~\ref{lema:soundness_rule}, we have $S_{n-1}$ is false under any interpretation, too. Similarly, $S_{n-1} , \ldots , S_1 , S_0$ are also false under any  interpretation. This completes the proof of the soundness of the resolution procedure.

We now prove the completeness of the resolution procedure, that means if $S_0$ is unstatisfialbe then $S_n$ contains the empty clause. We apply the \textit{semantic tree} method for two-valued logics to our linguistic propositional logic.

Let $S$ be a set containing exactly $n$ atoms $A_1,A_2,\ldots,A_n$. A semantic tree of $S$ is an $n$-level complete binary tree, each level corresponds to an atom. The left edge of each node at the level $i$ is assigned with the label $A_i\leq \mathsf{W}$, and the right edge of each node at the level $i$ is assigned with the label $A_i >\mathsf{W}$ (cf.Fig~\ref{fig:semantic-tree}).

\begin{figure}[h]
\centering
\begin{tikzpicture}[-, main node/.style={circle,fill=black!50,draw},minimum width=2pt]

  \node[main node] (1) {};
  \node[main node] (2) [below left of=1, node distance=3cm] {};
  \node[main node] (3) [below right of=1, node distance=3cm] {};
  \node[main node] (4) [below left of=2, node distance=2cm] {};
  \node[main node] (5) [below right of=2, node distance=2cm] {};
  \node[main node] (6) [below left of=3, node distance=2cm] {};
  \node[main node] (7) [below right of=3, node distance=2cm] {};
  \node (8) [below of=4, node distance=1cm] {};
  \node (9) [below of=5, node distance=1cm] {};
  \node (10) [below of=6, node distance=1cm] {};
  \node (11) [below of=7, node distance=1cm] {};

  \path[every node/.style={font=\sffamily\small}]
  	(1) edge node [left] {$A_1 \leq \mathsf{W}$} (2)
  		edge node [right] {$A_1 > \mathsf{W}$} (3)
  	(2) edge node [left] {$A_2 \leq \mathsf{W}$} (4)
  		edge node [right] {$A_2 > \mathsf{W}$} (5)
  	(3) edge node [left] {$A_2 \leq \mathsf{W}$} (6)
  		edge node [right] {$A_2 > \mathsf{W}$} (7)
    
	;
	\draw [dashed] (4) -- (8);
	\draw [dashed] (5) -- (9);
	\draw [dashed] (6) -- (10);
	\draw [dashed] (7) -- (11);
\end{tikzpicture}
\caption{Semantic tree}
\label{fig:semantic-tree}
\end{figure}
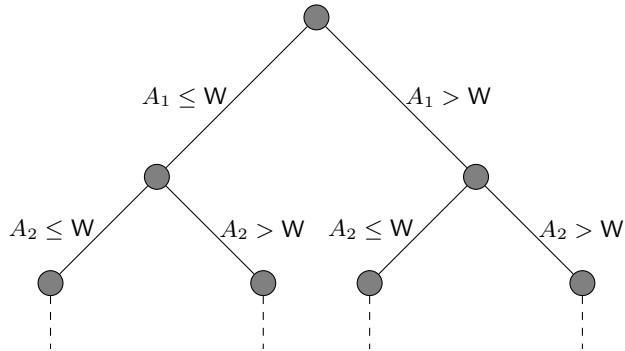

A set of clauses $S$ \emph{is failed at the node} $t$ of a semantic tree $T$ iff there exist an interpretation $I$ corresponding to a branch in $T$ which contains $t$, such that $S$ is false under $I$. A node $t$ is called a \emph{failure node} of $S$ iff $S$ fails at $t$ and does not fail at any node above $t$. A node $t$ in a semantic tree $T$ is called an \emph{inference node} iff both successor nodes of $t$
are failure nodes. If there are failure nodes for $S$ on every branch of the corresponding semantic tree $T$, removing all child nodes of each failure node, we receive a \emph{failure tree} $FT$.

 Assume that we have a failure tree $FT$. Because $FT$ has finite level, so there exists one (or more) leaf node on $FT$ at the highest level, let say this node is called $j$. Let $i$ be parent node of $j$. By definition of failure tree, $i$ cannot be failure node. Therefore, $i$ has another child node, named $k$ (Figure~\ref{fig:proof-inference}). If $k$ is a failure node then $i$ is inference node, the lemma is proved. If $k$ is not a failure node then it has two child nodes: $l,m$. Clearly $l,m$ are at higher level than $j$. This contradicts with the assumption that $j$ is at the highest level. Therefore $k$ is a failure node and $i$ is an inference node. 

\begin{figure}[h!]
\centering
\begin{tikzpicture}[-, main node/.style={circle,fill=white!25,draw,node distance=3cm},minimum width=3pt]

  \node[main node] (1) {i};
  \node[main node] (2) [below left of=1] {k};
  \node[main node] (3) [below right of=1] {j};
  \node[main node] (4) [below left of=2] {l};
  \node[main node] (5) [below right of=2] {m};

  \path[every node/.style={font=\sffamily\small}]
  	(1) edge node [left] {} (2)
  		edge node [right] {} (3)
  	(2) edge node [left] {} (4)
  		edge node [right] {} (5)
	;

\end{tikzpicture}
\caption{Inference node on failure tree}
\label{fig:proof-inference}
\end{figure}
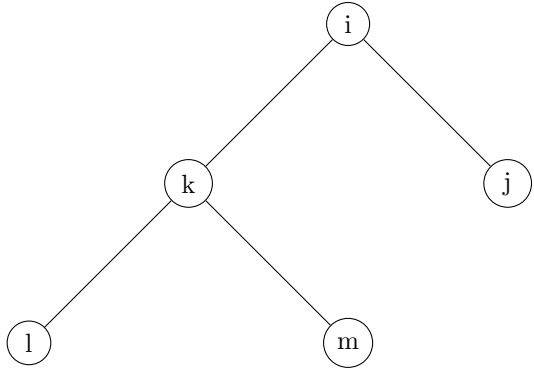

Let $FT_1$ (respectively $FT_2$) be a failure tree of the set of
clauses $S_1$ (respectively $S_2$). We denote $FT_1\supset FT_2$ iff there exists an inference node $i$ of $FT_1$ such that removing two successor nodes of $i$ on $FT_1$ we receive $FT_2$.

Because $S_0$ is unsatisfiable, there is a corresponding failure tree $FT$ and an inference node $i$ on $FT$ with two child nodes $j,k$. Assuming that the label of edge $i-j$ is $A \leq \mathsf{W}$ and the label of edge $i-k$ is $A > \mathsf{W}$. The interpretation corresponding to the branch contains the edge $i-j$ and terminating at $j$ makes $S_0$ satisfiable. So, there is at least one clause in $S_0$ containing the literal $A^{\alpha_1}$ where $\alpha_1 \leq \mathsf{W}$, let say $C_1$. Similarly, there exists at least one clause in $S_0$ containing the literal $A^{\alpha_2}$ where $\alpha_2 > \mathsf{W}$, we named it $C_2$. Applying the resolution rule~\ref{defi:resolution-rule}: $$\frac{C_1,C_2}{C_3}$$ $C_3$ does not contain atom $A$, so that $C_3$ is false under all interpretations containing $A$. Thus, failure tree $FT_1$ of clause set $S_1=S_0 \cup C_3$ does not contain node $j,k$, this means $FT\supset FT_1$.

By applying resolution procedure, there exist failure tree $FT_2,FT_3,\ldots$ of the sets of clauses $S_2,S_3,\ldots$ such that $FT\supset FT_1\supset FT_2\supset FT_3\supset\ldots$. Because there are only a finite number of nodes in $FT$, then exists some $n$ satisfying: $FT_n=\Box$ (i.e. $FT\supset FT_1\supset FT_2\supset FT_3\supset\ldots\supset FT_n=\Box$). Only the empty clause is false under the empty interpretation. This means that the set of clauses $S_n$ ($S_n$ corresponds to $FT_n$) contains the empty clause. This completes the proof of the theorem.
\end{proof}

\section{Proof of Lemma~\ref{lemma:RT2}}
\label{proof_lemmaRT2}
\textbf{Lemma~\ref{lemma:RT2}.} Consider the following resolution inferences:
$$ \frac{(A^a \lor B^{b_1},\alpha),(B^{b_2} \lor C^c,\beta)}{(A^a \lor C^c,\gamma)}$$
$$ \frac{(A^a \lor B^{b_1},\alpha),(B^{b_2} \lor C^c,\beta')}{(A^a \lor C^c,\gamma')}$$
Then, $\beta' > \beta$ implies $\gamma' \geq \gamma$.
\begin{proof}
We have 

$\gamma = \alpha \land \beta \land \neg (b_1 \land b_2) \land (b_1 \lor b_2)$

$\gamma' = \alpha \land \beta' \land \neg (b_1 \land b_2) \land (b_1 \lor b_2)$

Denote $\delta = min(\alpha, \neg (b_1 \land b_2),(b_1 \lor b_2))$ then $\gamma=min(\beta,\delta)$, $\gamma=min(\beta',\delta)$.

Hence, if $\beta' > \beta$ implies $\gamma' \geq \gamma$

The equality occurs when $\beta \ge min(\alpha, \neg (b_1 \land b_2),(b_1 \lor b_2))$.
\end{proof}

\section{Proof of Lemma~\ref{lemma:saturated}}
\label{proof_lemmasaturated}
\textbf{Lemma~\ref{lemma:saturated}. }
  Let $S_0, \ldots, S_i, \ldots$ be a fuzzy linguistic resolution
  $\alpha$-strategy derivation, and $S_\infty$ be the the limit of the derivation.  Then $S_\infty$ is saturated.
\begin{proof} 
  By contradiction assume that $S_\infty$ is not saturated. Then there
  must be a fuzzy linguistic resolution inference
  $$\frac{(C_1,\alpha_1), (C_2, \alpha_2)}{(C,\alpha)}$$ 
  where $(C_1,\alpha_1), (C_2, \alpha_2) \in S_\infty$, there is not
  any clause $(C,\alpha') \in S_\infty$ such that $\alpha \leq
  \alpha'$.  By definition of $\alpha$-strategy derivation, either
  $(C,\alpha)$ is in $S_\infty$ or there must be a clause
  $(C,\alpha'')$ in $S_i$ for some $i=0, 1, \ldots$ such that $\alpha
  \leq \alpha''$, this also means that $(C,\alpha)$ is removed from
  $S_j$ for some $j \geq i$. In both cases, we have a contradiction.
\end{proof}

\section{Proof of Theorem~\ref{theo:max_reli}}
\label{proof_maxreli}
\textbf{Theorem~\ref{theo:max_reli}. }
  Let $S_0, \ldots, S_i, \ldots$ be a fuzzy linguistic resolution
  $\alpha$-strategy derivation, and $S_\infty$ be the the limit of the
  derivation.  Then for each clause $(C, \alpha)$ in $S_\infty$, there
  is not any other resolution proof of the clause $(C, \alpha')$ from
  $S_0$ such that $\alpha' > \alpha$.

\begin{proof}
  By contradiction, suppose that for some clause $(C,\alpha)$ in
  $S_\infty$, there exists a resolution proof of $(C,\alpha')$ from
  $S_0$ such that $\alpha' > \alpha$. Let $(C_1, \alpha_1)$ and $(C_2,
  \alpha_2)$ be the two parents of $(C, \alpha')$ in such a resolution
  proof of $(C, \alpha')$. We have that $(C_1, \alpha_1)$ and $(C_2,
  \alpha_2)$ cannot be both in $S_\infty$ because otherwise an
  inference with these two clauses as premisses would give
  $(C,\alpha')$ in $S_\infty$. Without lost of generality, we can
  assume that $(C_1, \alpha_1)$ is not in $S_\infty$. Obviously, the
  resolution proof of $(C_1, \alpha_1)$ from $S_0$ can not be in
  $S_\infty$. That also means there is a clause $(C_1, \alpha_1')$ in $S_\infty$ such that there is not any other resolution proof of the clause $(C_1,\alpha_1'')$, where $\alpha_1'' > \alpha_1'$. By Lemma \ref{lemma:saturated}, $S_\infty$ is satutared. According to Lemma
  \ref{lemma:RT2}, the inference with premisses $(C_1,\alpha_1')$ and
  $(C_2,\alpha_2)$ gives us the conclusion $(C,\alpha'')$, with
  $\alpha'' > \alpha'$. This contradicts with the fact that $S_\infty$ is satutared. This completes the proof of the theorem.
\end{proof}

\end{appendix}
\end{document}